\documentclass[12pt]{iopart}

\usepackage{setstack}
\usepackage{amssymb}
\usepackage{theorem}
\usepackage{citesort}

\newcommand\pdf{\phi}
\newcommand\Var{\mathrm{Var}}

\newcommand\Set[1]{\{ #1 \}}

\newcommand\Oh{\mathcal{O}}
\newcommand\filledSquare{\rule{7pt}{7pt}}
\newcommand\T{\mathcal{T}}
\newcommand\U{\mathcal{U}}

\newcommand\E{\mathbb{E}}

\renewcommand\epsilon\varepsilon

\newcommand{\suppress}[1]{}

\newcommand{\indentedline}[1]{\par\noindent\hspace{#1 em}}
\newcommand{\komment}[1]{{\footnotesize \tt \ \% #1}}

\theorembodyfont{\slshape}
\newtheorem{theorem}{Theorem}
\newtheorem{lemma}[theorem]{Lemma}
\newtheorem{corollary}[theorem]{Corollary}
{\theorembodyfont{\rmfamily} \newtheorem{remark}{Remark}}
\newenvironment{proof}{\par\noindent{\bf Proof: }}{\nopagebreak\hfill\filledSquare\medskip}
{\theorembodyfont{\rmfamily} \newtheorem{algorithm}{Algorithm}}

\newcommand{\iid}[1]{{\it i.i.d.}#1}
\newcommand{\ie}[1]{{\it i.e.}#1}

\newcommand{\strongMC}[1]{{\sc BoundedMC}#1}
\newcommand{\aggregate}[1]{{\sc BTT}#1}
\newcommand{\recAggregate}[1]{{\sc RecBTT}#1}
\newcommand{\basketAggregate}[1]{{\sc BasketBTT}#1}
\newcommand{\merge}[1]{{\sc Merge}#1}
\newcommand{\estimate}[1]{{\sc Estimate}#1}

\newcommand{\val}{\mathrm{.value}}
\newcommand{\mass}{\mathrm{.mass}}
\newcommand{\range}{\mathrm{range}}

\newcommand\binom[2]{{#1\choose#2}}
\renewcommand\text[1]{\mbox{#1}}

\begin{document}

\title[Fast Pricing of European Asian Options]{Fast Pricing of European Asian Options with Provable Accuracy: Single-stock and Basket Options}
\author{Karhan Akcoglu\dag\footnote[4]{Supported in part by NSF Grant CCR-9896165.},
Ming-Yang Kao\ddag\footnote[5]{Supported in part by NSF Grants CCR-9531028 and CCR-9988376.  Part of this work was performed while this author was visiting Department of Computer Science, Yale University.}, and 
Shuba Raghavan\S}

\address{\dag \ Department of Computer Science, Yale University, New Haven, CT 06520, USA.}

\address{\ddag \ Department of Electrical Engineering and
Computer Science, Tufts University, Medford, MA 02155, USA.}

\address{\S \ Department of Economics, Yale University, New Haven, CT 06520, USA.}

\eads{\mailto{karhan.akcoglu@yale.edu}, \mailto{kao@eecs.tufts.edu}, \mailto{kao@cs.yale.edu}}

\begin{abstract}
This paper develops three polynomial-time pricing techniques for
Eu\-ro\-pe\-an Asian options with provably small errors, where the stock
prices follow binomial trees or trees of higher-degree.  The first
technique is the first known Monte Carlo algorithm with analytical
error bounds suitable for pricing single-stock options with meaningful
confidence and speed.  The second technique is a general recursive
bucketing-based scheme that can use the Aingworth-Motwani-Oldham
aggregation algorithm, Monte-Carlo simulation and possibly others as
the base-case subroutine. This scheme enables robust trade-offs between accuracy and time over subtrees of different sizes. For long-term options or high-frequency price averaging, it can price single-stock options with smaller errors in less time than the base-case algorithms themselves. The third technique combines Fast Fourier Transform with bucketing-based schemes for pricing basket
options. This technique takes polynomial time in the number of days and the number of stocks, and does not add any errors
to those already incurred in the companion bucketing scheme.  This
technique assumes that the price of each underlying stock moves
independently.
\end{abstract}

\maketitle

\section{Introduction}
A \emph{call} (respectively, \emph{put}) \emph{option} is a contract
assigning its \emph{holder} the right, but not the obligation, to buy
(respectively, sell) a security at some future time for a specified
{\emph{strike price}} $X$ \cite{Hull:2000:OFO}. If the
holder \emph{exercises} her right, the other party in the contract,
the {\emph{writer}}, is obligated to assume the opposite side of the
transaction. In exchange for this right, the holder pays the writer an
{\emph{option price}} $P$. The security in this contract can
be any financial asset; for the purpose of this paper, we restrict it
to a single stock or a portfolio of stocks.  An option in the
latter case is commonly called a {\it basket} option; for clarity, we
call an option in the former case a {\it single-stock} option.  

Options are popular financial instruments for a variety of trading
strategies. For example, options can be used to hedge risk. As
protection from a potential price fall in a stock price, one can purchase a
put on the stock, thereby locking in a minimum sell price.  On the
other hand, options can provide additional income for stockholders who
write calls on their holdings; of course, this strategy carries the
risk of being forced to sell the stock should the calls be exercised.

An option is valid until its \emph{expiry date}. For a
{\emph{European}} option, the holder may exercise it only on the
expiry date. For an \emph{American} option, the holder may exercise it
on any date up to and including the expiry date.  The
\emph{payoff} of an option is the amount of money its holder makes
on the contract.  A European call is worth exercising if and only if
$S \ge X$, where $S$ is the stock price on the expiry date.  The payoff of the call is $(S-X)^+=\max(S-X, 0)$.  For an American
call, $S$ is set to the stock price at the exercise time.  An {\it
Asian} option comes in European and American flavors, depending on
when it may be exercised.  For an European Asian call, if $A$ is the
average stock price over the entire life of the contract up to the
expiry date, the payoff is $(A-X)^+$.  For an American Asian
call, $A$ is set to the average stock price up to the exercise
date. The payoffs of puts can be symmetrically defined.

The price $P$ of an option is the discounted expected value of the payoff with an appropriate martingale measure. Because of the popularity of options, pricing
techniques for computing $P$ have been extensively researched
\cite{Baxter:1996:FCI,Neftci:1996:IMF,Wilmott:1995:MFD,Chalasani:1999:AOP,Chalasani:1999:AAE,Hull:2000:OFO}. Generally, it is more difficult to price a basket option
than a single-stock option.  To compute $P$, the price movement of
each individual stock needs to be modeled. Typically, it is modeled as
Brownian motion with drift. Using a stochastic differential equation,
$P$ can then be computed via a closed-form solution to the
equation. When a closed-form solution is not known, various approaches
are used to find an approximate solution. One class of approaches
involves approximating the solution using numerical methods. Other
approaches approximate the Brownian motion model with a discrete
model, and use this model to approximate $P$. One such discrete model
is the
\emph{binomial tree} model, due to Cox, Ross, and
Rubenstein~\cite{Cox:1979:OPS}; see Section~\ref{section-definitions} for the
definition of the model.

This paper develops three polynomial-time pricing techniques with
provably small errors.  The remaining discussion makes the following
assumptions:
\begin{enumerate}
\item 
The option in question is an European Asian single-stock or basket
call.
\item 
Our task is to price the call at the start of its contract life.
\item 
The price of each underlying stock follows the binomial tree model.
\item In the case of a basket call, 
the price of each underlying stock moves independently.
\end{enumerate}
Our results generalize easily for puts, for a later time point than
the start of the contract life, and for trees with higher degrees than
two. The cases of American options and of interdependent stocks remain
open.

Monte Carlo simulation has been commonly used in the financial
community.  Despite this popularity, most reported results on error
bounds are experimental or heuristic
\cite{Robert:1999:MCS,Aingworth:2000:AAA}.  Our first technique is the
first known Monte Carlo algorithm that has analytical error bounds
suitable for pricing a European Asian call with meaningful confidence
and speed. As shown by Theorem~\ref{theorem-algorithm}, the number of simulations
required is polynomial in (P1) the logarithm of the inverse of the
error probability and (P2) the inverse of the price error relative to
the strike price but is exponential in (E1) the square root of the number of 
underlying stocks and (E2) the volatility of these stocks over the
call's life. In particular, the algorithm is reasonably fast and
accurate for a single-stock European Asian call with reasonable
volatility.

Monte Carlo simulation is a randomized technique, and thus there is
always a nonzero probability that the price obtained by
polynomial-time Monte Carlo simulation is not accurate enough.  The
aggregation algorithm of Aingworth, Motwani, and Oldham (AMO)
\cite{Aingworth:2000:AAA} is the first polynomial-time algorithm for pricing
single-stock European Asian calls and other path-dependent options
with guaranteed worst-case price errors.  The AMO algorithm is based
on a simple yet powerful idea called {\it bucketing}.  Our second technique is
a general recursive bucketing-based scheme that can use the AMO
algorithm, Monte-Carlo simulation, and possibly others as the
base-case subroutine. This scheme enables robust trade-offs between
accuracy and time over subtrees of different sizes. For long-term options or high-frequency price averaging, it can price single-stock European Asian calls with smaller error bounds in less time than the base-case algorithms themselves.
In particular, as
implied by Theorem~\ref{theorem-recAggregate}, given the same runtime, this recursive scheme prices more accurately than the AMO algorithm; similarly, given the same accuracy, the scheme
runs faster than the AMO algorithm.

This recursive scheme works for calls written on a single stock.  Our
third technique combines Fast Fourier Transform (FFT)
\cite{Cormen:1999:IA} and bucketing-based schemes to price basket
calls and is applicable to European Asian calls as well as others. As
shown in Theorem~\ref{theorem-basket}, this technique takes polynomial
time in the number of days and the number of stocks, and does not add
any errors to those already incurred in the companion bucketing
schemes.

The remainder of this paper is organized as follows.
Section~\ref{section-definitions} reviews the binomial tree model and
basic definitions.  Section~\ref{section-MC} describes the new Monte
Carlo algorithm.  Section~\ref{section-aggregation} details the
recursive scheme.  Section~\ref{section-basket} gives the FFT-based
technique for pricing basket calls.  Section~\ref{section-open} concludes
the paper with directions for further research.

\section{The Binomial Tree Model}
\label{section-definitions}
A \emph{binomial} tree $\T$ is a \emph{recombinant} binary tree. If $n$ is the depth of $\T$, $\T$ has $t+1$ nodes at depth $t$, for $0\le t\le n$. For $0\le i \le t$, let $\T[t, i]$ (or simply $[t,i]$ if $\T$ is obvious from context) be the $i$-th node at level $t$ of $\T$. For $t > 0$, $\T[t, 0]$ and $\T[t, t]$ have one parent each, $\T[t-1, 0]$ and $\T[t-1, t-1]$ respectively. For $0 < i < t$, $\T[t, i]$ has two parents, $\T[t-1, i-1]$ and $\T[t-1, i]$. The number of nodes in $\T$ is $\frac{(n+1)(n+2)}{2}$.

Given a stock in the binomial tree model, the stock price is assumed to follow a geometric random walk through $\T$. Time is divided into $n$ equal periods, with the root $\T[0,0]$ corresponding to time $t=0$, when the option is priced, and the leaves $\T[n, \ \cdot\ ]$ corresponding to time $t = n$, the expiry date of the option. Let $s(\T[t,i])$ (or simply $s(t, i)$) be the stock price at node $\T[t,i]$.  At each time step, the stock price $s(t, i)$ rises to $s(t+1, i+1) = u\cdot s(t, i)$---an \emph{uptick}---with probability $p$ or falls to $s(t+1, i) = d\cdot s(t, i)$---a \emph{downtick}---with probability $q = 1-p$. Letting $r$ denote the \emph{risk-free interest rate}, the parameters $u$ and $d$ satisfy $0 < d \le 1 + r \le u$ and are typically taken to be $u = \frac{1}{d} = e^{\sigma/\sqrt{n}}$, where $\sigma$ is the $n$-period \emph{volatility}, or standard deviation, of the stock price \cite{Hull:2000:OFO}. Although the probability $p$ of an uptick is not known in general, for the purposes of pricing options, we can use the \emph{risk-neutral probability model} \cite{Hull:2000:OFO}, which states that $p = \frac{(1+r) - d}{u - d}$, where $r$ is the risk-free interest rate for one period. This makes the expected return on the stock over one period, $pu + (1-p)d$, equal to the risk-free return, $1+r$.

Let $\Omega$ be the sample space of paths $\omega = (\omega_1, \ldots, \omega_{n})$ down $\T$, where each $\omega_t \in \Set{-1, 1}$, with $-1$ corresponding to a downtick and $1$ corresponding to an uptick. Given $\omega\in\Omega$ and $0\le t\le n$, let $\T[t, \omega]$ be the unique node at level $t$ that $\omega$ passes through. Similar to the notation introduced above, we let $s(\T[t, \omega]) = s(t, \omega)$ be the price at node $\T[t, \omega]$.

We define the random variables $Y_1, \ldots, Y_{n}$ on $\Omega$ by $ Y_t(\omega) = \omega_t$, the $t$-th component of $\omega$. We define the probability measure $\Pi$ on $\Omega$ to be the unique measure for which the random variables $Y_1, \ldots, Y_{n}$ are independent, identically distributed (\iid{}) with $P(Y_i = 1) = p$ and $P(Y_i = -1) = q$. We start with an initial fixed stock price $S_0 = s(\T[0,0])$. For $1 \le t \le n$, the stock price $S_t$ is a random variable defined by $S_t = S_0 u^{\sum_{i=1}^t Y_i}$. From the structure of the binomial tree, we have $\Pr(S_t = s(t, i)) = \binom{t}{i} p^i(1-p)^{t-i}$, where $0\le i\le t$. The \emph{running total} of the stock prices is defined by $T_t = \sum_{i=0}^t S_i$ and the \emph{running average} is $A_t = T_t/(t+1)$. For $\omega\in\Omega$, we let $S_t(\omega) = S_0 u^{\sum_{i=1}^t Y_i(\omega)}$, $T_t(\omega) = \sum_{i=0}^t S_i(\omega)$, and $A_t(\omega) = T_t(\omega)/(t+1)$.

Recall that $X$ is the strike price of a European Asian call. Using the above notation, the price of this call is
\[
\fl
\E\big((A_n - X)^+\big) = \frac{1}{n+1}\E\big((T_n - (n+1)X)^+\big) = \frac{1}{n+1}\E\big(\max(T_n - (n+1)X, 0)\big).
\]
For large $n$, it is not known if this quantity can be computed exactly because the stock price can follow exponentially many paths down the tree. Below, we show how to estimate $\E\big((T_n - (n+1)X)^+\big)$, from which the price of the option can be easily computed.

\section{A New Monte Carlo Algorithm}
\label{section-MC}

Monte Carlo simulation methods for asset pricing were introduced to finance by 
Boyle \cite{Boyle:1977:OMC}. They are very popular in pricing
complex instruments, particularly path-dependent 
European-style options. These methods involve randomly sampling paths $\omega\in\Omega$ according to the distribution $\Pi$ and computing the payoff $(A_n(\omega) - X)^+$ on each sample. Suppose $N$ samples $\omega^1, \ldots, \omega^N$ are taken from $\Omega$. The price estimate of the call is
\[
\mu = \frac{1}{N}\sum_{i=1}^N (A_n(\omega^i) -X)^+.
\]
The accuracy of this estimate depends on the number of simulations $N$ and the variance $\tau^2$ of the payoff: the error bound typically guaranteed by Monte Carlo methods is $\Oh(\tau/\sqrt{N})$. Generally $\tau^2$ is not known, and itself must be estimated to determine the error bound of $\mu$.

A number of techniques are used to reduce the error of $\mu$. For example, \cite{Clewlow:1994:SCC,Vazquez-Abad:1998:ASP} use the control variate technique, which ties the price of the option to the price of another instrument (the control variate) for which an analytically tractable solution is known. The simulation then estimates the difference between the option price and the control variate, which can be determined with greater accuracy than the option price itself. The antithetic variate method \cite{Costantini:1999:VRA,Hammersley:1956:NMC} follows not only randomly sampled paths $\omega$ down the binomial tree, but also the ``mirror images'' of each $\omega$. None of these techniques have known analytical error bounds for $\mu$.

Below we use concentration of measure results in conjunction with Monte Carlo simulation to estimate the price of a European Asian call and derive analytical error bounds for this estimate. The error bounds are in terms of the strike price $X$ of the option and the maximum volatility $\sigma_{\max{}}$ of the underlying stocks.

\subsection{Analytical Error Bounds for the Single-stock Case}

Let $C=e^{\E(\ln T_n)}$. In this section, we show that if $(n+1)X/C$ is ``small'', $\E\big((T_n-(n+1)X)^+\big)$ is close to $\E(T_n-(n+1)X) = \E(T_n)-(n+1)X$ (Theorem~\ref{theorem-bounds}(1)). The option is \emph{deep-in-the-money} and will probably be exercised. Since a closed-form formula exists for $\E(T_n)$, as described in Lemma~\ref{lemma-expected}, $\E(T_n-(n+1)X)$ can be computed exactly, and our algorithm uses it as our estimate for $\E\big((T_n-(n+1)X)^+\big)$. On the other hand, if $(n+1)X/C$ is not small, the variance of $(T_n-(n+1)X)^+$ can be bounded from above (Theorem~\ref{theorem-bounds}(2)) and our algorithm estimates its expectation with bounded error using Monte Carlo simulation.

We first give some theoretical results, then show how these results can be used in our Monte Carlo algorithm, \strongMC{}. We begin by finding bounds $A$ and $B$ such that $T_n \in [A, B]$ with high probability. Our main tool for this is an inequality arising from the theory of martingales, usually known as Azuma's Inequality \cite{Azuma:1967:WSC,Motwani:1995:RA}. We use the inequality in a form taken from Frieze \cite{Frieze:1991:LLM}. Suppose we have a random variable $U = U(V)$, where $V = (V_1, \ldots, V_\eta)$ and, for $1\le i\le \eta$, $V_i$ is chosen independently from the probability space $\Omega_i$. For $V,W \in \Omega = \Omega_1\times\cdots\times\Omega_\eta$, write $V\approxeq W$ if $V$ and $W$ differ in exactly one component.

\begin{theorem}[Azuma's Inequality]
\label{theorem-azuma}
Suppose that $V\approxeq W$ implies $|U(V) - U(W)| \le a$, for some constant $a > 0$. Then, for any real $b\ge 0$, 
\[
\Pr(|U - \E(U)|\ge b)\le 2e^{-\frac{2b^2}{a^2\eta}}.
\]
\end{theorem}

\begin{lemma}
\label{lemma-azuma}
Let $C = e^{\E(\ln T_n)}$. For any $\lambda > 0$, we have 
\[
\Pr\big(T_n \le Ce^{-\sigma\lambda} \text{ or } T_n \ge Ce^{\sigma\lambda}\big) \le 2e^{-\lambda^2/2},
\]
where $\sigma$ is the volatility of the stock.
\end{lemma}

\begin{proof}
We apply Azuma's Inequality to the random variable $U = \ln T_n$. Let $\omega_X, \omega_Y \in \Omega$ be two paths that differ in a single component, say on day $t$. Assume without loss of generality that $\omega_X$ has an uptick on day $t$ while $\omega_Y$ has a downtick. Then $\frac{T_n(\omega_X)}{T_n(\omega_Y)} \le u^2$ and $|\ln\frac{T_n(\omega_X)}{T_n(\omega_Y)}| = |\ln T_n(\omega_X) - \ln T_n(\omega_Y)| \le 2\ln u = 2\frac{\sigma}{\sqrt{n}}$.

Applying Azuma's Inequality with $a = 2\frac{\sigma}{\sqrt{n}}$ and $b = \sigma\lambda$, we have
\[
\Pr(|\ln T_n - \E(\ln T_n)| \ge \sigma\lambda) \le 2e^{-\lambda^2 / 2}.
\]
The claimed result follows from the fact that
\begin{eqnarray*}
\fl
|\ln T_n - \E(\ln T_n)| \ge \sigma\lambda
&\Leftrightarrow& \ln T_n - \E(\ln T_n) \ge \sigma\lambda \text{ or } \ln T_n - \E(\ln T_n) \le - \sigma\lambda \\
&\Leftrightarrow& e^{\ln T_n}/C \ge e^{\sigma\lambda} \text{ or } e^{\ln T_n}/C \le e^{- \sigma\lambda}.
\end{eqnarray*}
\end{proof}

Now, fix $\varepsilon > 0$ and choose $\lambda_0 = \sqrt{2\ln\frac{2}{\varepsilon}}$. Then, by Lemma~\ref{lemma-azuma},
\[
\Pr\big(T_n \le Ce^{-\sigma\lambda} \text{ or } T_n \ge Ce^{\sigma\lambda}\big) \le \varepsilon.
\]
Theorem~\ref{theorem-bounds}(1) says that if $(n+1)X < Ce^{-\sigma \lambda_0}$, then $\E\big((T_n-(n+1)X)^+\big)$ is close to $\E(T_n-(n+1)X)$. Otherwise, Theorem~\ref{theorem-bounds}(2) says that $\Var\big((T_n -(n+1)X)^+ \big)$ is bounded.

\begin{theorem}
\label{theorem-bounds}
Let $C = e^{\E(\ln T_n)}$.
\begin{enumerate}
\item
If $(n+1)X < Ce^{-\sigma \lambda_0}$, then
\[
\left|\E \big((T_n -(n+1)X)^+ \big) - \E\big( T_n -(n+1)X \big) \right|  \leq \epsilon (n+1)X.
\]
\item
If $(n+1)X \ge Ce^{-\sigma\lambda_0}$, then
\[
\Var\big( (T_n -(n+1)X)^+ \big) \leq
  (n+1)^2 X^2 e^{4 \sigma \lambda_0} \frac{1 + 2 \sigma \epsilon}{\lambda_0 - 2\sigma}.
\]
\end{enumerate}
\end{theorem}

\begin{proof} Let $\pdf(t)$ denote the probability density function of $T_n$.

Statement 1.  
Note first that
\begin{eqnarray*}
\fl
\E\big((T_n-(n+1)X)^+\big) &=& \int_0^\infty (t-(n+1)X)^+ \pdf(t) dt
\quad=\quad \int_{(n+1)X}^\infty (t-(n+1)X) \pdf(t) dt \\
&=& \E(T_n-(n+1)X) - \int_0^{(n+1)X} (t-(n+1)X) \pdf (t) dt.
\end{eqnarray*}
Then
\begin{eqnarray*}
\fl
\big|\E\big((T_n-(n+1)X)^+\big) - \E(T_n-(n+1)X) \big| \\
\lo=  \left| \int_0^{(n+1)X} (t-(n+1)X) \pdf (t) dt \right|
\quad\le\quad (n+1)X  \int_0^{(n+1)X} \pdf (t) dt \\
\lo= (n+1)X \Pr\big(T_n \le (n+1)X\big)
\quad\le\quad (n+1)X \Pr\big(T_n \le Ce^{-\sigma\lambda_0}\big) \\
\lo\leq  \epsilon (n+1)X,
\end{eqnarray*}
where the second-last inequality follows from the assumption that 
$(n+1)X < Ce^{- \sigma \lambda_0}$ and the last inequality follows from
Lemma~\ref{lemma-azuma} and our choice of $\lambda_0$.

Statement 2.  
Since $(n+1)X \ge  Ce^{-\sigma\lambda_0}$,
\begin{eqnarray*}
\fl
\Var\big( (T_n-(n+1)X)^+ \big) \ \le\  \E\Big(\big( (T_n -(n+1)X)^+ \big)^2\Big) \ =\  \int_{(n+1)X}^\infty (t-(n+1)X)^2 \pdf(t) dt \\
\lo=  2\int_{(n+1)X}^\infty (t-(n+1)X) \Pr(T_n \geq t) dt \\
\lo\le 2 \int_{Ce^{-\sigma\lambda_0}}^\infty 
           (t - C e^{-\sigma \lambda_0}) \Pr(T_n \ge t)dt \\
\lo= 2\int_{Ce^{-\sigma \lambda_0}}^{Ce^{\sigma \lambda_0}} 
        (t-Ce^{-\sigma\lambda_0}) \Pr(T_n \geq t) dt +
    2 \int_{Ce^{\sigma\lambda_0}}^\infty 
        (t-Ce^{-\sigma \lambda_0}) \Pr(T_n \geq t) dt \\
\lo\le  C^2 e^{2\sigma\lambda_0} + 
          4 C^2 \sigma \int_{\lambda_0}^\infty 
                    e^{ 2\sigma \lambda} e^{-\lambda^2/ 2} d\lambda \\
   \quad\quad \text{(substitute $t=Ce^{\sigma\lambda}$ in the second term)} \\
\lo= C^2 e^{2\sigma\lambda_0} + 4 C^2 \sigma e^{2 \sigma^2} \int_{\lambda_0}^\infty 
                    e^{-\frac{1}{2}(\lambda - 2\sigma)^2 }d\lambda \\
\lo\le  C^2 e^{2\sigma\lambda_0} + 4 C^2 \sigma
          \frac{e^{2\sigma \lambda_0} e^{-\lambda_0^2/2}}{ \lambda_0 - 2\sigma} 
   \quad \text{(from the inequality ${\textstyle \int_{t}^\infty e^{-\frac{1}{2} x^2}dx \le \frac{1}{t}e^{-\frac{1}{2}t^2}}$)}   \\
\lo\le (n+1)^2 X^2 e^{4 \sigma \lambda_0}\left(1 + \frac{2\sigma\varepsilon}{\lambda_0 - 2\sigma} \right) \\
      \quad\quad\hbox{(since $(n+1)Xe^{2\sigma\lambda_0} \ge Ce^{\sigma\lambda_0}$ and $\epsilon = 2e^{-\lambda_0^2/2}$)} \\
\lo\le  (n+1)^2 X^2 e^{4 \sigma \lambda_0} \frac{1 + 2 \sigma \epsilon}{\lambda_0 - 2\sigma}.
\end{eqnarray*}
\end{proof}

\subsection{The \strongMC{} Algorithm}

We next use these results in our algorithm. One approach would be to estimate $C = e^{\ln T_n}$ to determine whether we should apply Theorem~\ref{theorem-bounds}(1) or \ref{theorem-bounds}(2). Our algorithm takes a more direct approach. We begin by selecting $N$ samples $\omega^1,\ldots,\omega^N \in \Omega$ and computing $T_n(\omega^1), \ldots, T_n(\omega^N)$. For $1\le i\le N$, define the random variable $Z_i$ as
\[
Z_i = 
\left\{
\begin{array}{ll}
1&\text{if $T_n(\omega^i) \le (n+1)X$,} \\
0&\text{otherwise.}
\end{array}
\right.
\]
Let $Z = \sum_{i=1}^N Z_i$.

\begin{lemma}
\label{lemma-chernoff}
Let $0 < \delta < 1$ be given. With $N = \Theta(\log\frac{1}{\delta})$ trials, the following statements hold.
\begin{enumerate}
\item
If $\frac{Z}{N} \le 2\varepsilon$, then $\Pr\big(T_n \le (n+1)X  \big) \le 4\varepsilon$ with probability $1 - \delta$.
\item
If $\frac{Z}{N} > 2\varepsilon$, then $\Pr\big(T_n \le (n+1)X  \big) \ge \varepsilon$ with probability $1 - \delta$.
\end{enumerate}
\end{lemma}

\begin{proof}
This follows from the Chernoff bound \cite[Theorems 4.1 and 4.2]{Motwani:1995:RA}.
\end{proof}

\begin{theorem}
\label{theorem-algorithm}
Let $0 < \delta < 1$ be given. With $N = \Theta(\log\frac{1}{\delta} + \frac{1}{\varepsilon^2}e^{4\sigma\lambda_0}\frac{1+ 2\sigma\varepsilon}{\lambda_0 - 2\sigma})$ trials, the following statements hold.
\begin{enumerate}
\item
If $\frac{Z}{N} \le 2\varepsilon$, then, with probability $1 - \delta$, $\E(T_n - (n+1)X)$ estimates $\E\big((T_n - (n+1)X)^+\big)$ with error at most $4\varepsilon(n+1)X$. Correspondingly, the price of the call is estimated with error at most $4\varepsilon X$.
\item
If $\frac{Z}{N} > 2\varepsilon$, then, with probability $1 - \delta$, $\frac{1}{N}\sum_{i=1}^N \big(T_n(\omega^i) - (n+1)X\big)^+$ estimates $\E\big((T_n - (n+1)X)^+\big)$ with standard deviation at most $\varepsilon(n+1)X$. Correspondingly, the price of the call is estimated with standard deviation at most $\varepsilon X$.
\end{enumerate}
\end{theorem}

\begin{proof} The statements are proved as follows.

Statement 1.  
With probability $1 - \delta$, we have
\begin{eqnarray*}
\fl
|\E\big((T_n - (n+1)X)^+\big) - \E(T_n - (n+1)X)| \le (n+1)X\Pr(T_n \le (n+1)X)\\ \lo\le 4\varepsilon (n+1)X,
\end{eqnarray*}
where the first inequality follows from the proof of Theorem~\ref{theorem-bounds}(1) and the second inequality follows from Lemma~\ref{lemma-chernoff}(1). Since the price of the call is just $\frac{1}{n+1} \E\big((T_n - (n+1)X)^+\big)$, the error in our estimate for the price is at most $4\varepsilon X$. We give a closed-form formula for $\E(T_n - (n+1)X) = \E(T_n) - (n+1)X$ in Lemma~\ref{lemma-expected} below. 

Statement 2.  
If $\frac{Z}{N} > 2\varepsilon$, then by Lemma~\ref{lemma-chernoff}(2), with probability $1 - \delta$, $\Pr(T_n \le (n+1)X) \ge \varepsilon$. This implies $(n+1)X \ge Ce^{-\sigma\lambda_0}$. Otherwise, if $(n+1)X < Ce^{-\sigma\lambda_0}$, we would have
\[
\varepsilon \le \Pr(T_n \le (n+1)X) \le \Pr(T_n \le Ce^{-\sigma\lambda_0}),
\]
which contradicts our choice of $\lambda_0$ above. Since $(n+1)X \ge Ce^{-\sigma\lambda_0}$, we can apply Theorem~\ref{theorem-bounds}(2). The variance of our estimate $\frac{1}{N}\sum_{i=1}^N (T_n(\omega^i) - (n+1)X)^+$ for $\E\big((T_n - (n+1)X)^+\big)$ is $\frac{1}{N}\Var\big( (T_n -(n+1)X)^+ \big)$. Since
\[
\Var\left( (T_n -(n+1)X)^+ \right) \leq
  (n+1)^2 X^2 e^{4 \sigma \lambda_0} \frac{1 + 2 \sigma \epsilon}{\lambda_0 - 2\sigma},
\]
the variance of our estimate is at most $\varepsilon^2 (n+1)^2X^2$ for $N \ge \frac{1}{\varepsilon^2}e^{4\sigma\lambda_0}\frac{1+ 2\sigma\varepsilon}{\lambda_0 - 2\sigma}$ and its standard deviation is $\varepsilon (n+1) X$. Since the price of the call is just $\frac{1}{n+1} \E\big((T_n - (n+1)X)^+\big)$, the expected error in our estimate for the payoff is at most $\varepsilon X$. 
\end{proof}

We now summarize our algorithm.

\begin{algorithm} \strongMC{($\delta$, $\varepsilon$)}
\indentedline{0} generate $N = \Theta(\log\frac{1}{\delta} + \frac{1}{\varepsilon^2}e^{4\sigma\lambda_0}\frac{1+ 2\sigma\varepsilon}{\lambda_0 - 2\sigma})$ paths, $\omega^1, \ldots, \omega^N$;
\indentedline{0} let $Z$ be the number of paths $\omega^i$ such that $T_n(\omega^i) \le (n+1)X$;
\indentedline{0} if $Z/N \le 2\varepsilon$
\indentedline{1} return $\frac{1}{n+1}\E(T_n - (n+1)X) = \frac{1}{n+1}(\E(T_n) - (n+1)X)$;
\indentedline{0} else
\indentedline{1} return $\frac{1}{N}\sum_{i=1}^N (A_n(\omega^i) - X)^+$.
\end{algorithm}

For the completeness of Theorem~\ref{theorem-algorithm}, we conclude this section with a formula for $\E(T_n)$ \cite{Aingworth:2000:AAA}.

\begin{lemma}
\label{lemma-expected}
Let $S_0$ be the initial stock price and let $r$ be the interest rate. Then
\[
\E(T_n) = 
\left\{
\begin{array}{ll}
(n+1)S_0, &\text{if $r=0$,} \\[5pt]
\frac{(1+r)^{n+1} - 1}{r}S_0, &\text{if $r>0$.}
\end{array}
\right.
\]
\end{lemma}

\begin{proof}
Recall that $s(t, \ell)$ is the stock price at node $\T[t, \ell]$,
 $u$ and $d$ are the up\-tick/down\-tick factors, and $p$ and $q$ are the risk neutral probabilities. We have
\begin{eqnarray*}
\E(T_n) &=& \sum_{\text{level $0\le t\le n$}}\ \ \sum_{\text{node $0\le \ell\le t$}} s(t, \ell)\cdot\Pr(S_t = s(t, \ell)) 
\\
&=& \sum_{t=0}^n\sum_{\ell =0}^t S_0 u^\ell d^{t-\ell } \cdot \binom{t}{\ell } p^{\ell }(1-p)^{t-\ell } 
\\
&=& S_0 \sum_{t=0}^n \sum_{\ell =0}^t \binom{t}{\ell } (pu)^\ell ((1-p)d)^{t-\ell } 
\\
&=& S_0 \sum_{t=0}^n (pu + (1-p)d)^t \quad\text{(using the binomial theorem)}
\\
&=& S_0 \sum_{t=0}^n (1+r)^t \quad\text{(assuming risk-neutral rate of return)}
\\
&=&
\left\{
\begin{array}{ll}
(n+1)S_0, &\text{if $r=0$,} \\[5pt]
\frac{(1+r)^{n+1} - 1}{r}S_0, &\text{if $r>0$,}
\end{array}
\right.
\end{eqnarray*}
as claimed.
\end{proof}

\subsection{Pricing Basket Options using \strongMC{}}

The results derived above are applicable for European Asian basket calls as well. Suppose we have a basket of $m$ stocks. Let $T_n$ denote the running total of the stocks in the basket up to day $n$, let $\Omega^i$ be the sample space of paths the binomial tree corresponding to stock $i$ and let $\Pi^i$ be the corresponding measure on $\Omega^i$. Let $\Omega = \Omega^1 \times \cdots \times \Omega^m$ and $\Pi = \Pi^1 \times\cdots\times \Pi^m$. Lemma~\ref{lemma-azuma} is again applicable, but this time our random variable $U = \ln T_n$ has $mn$ components. Letting $\sigma_{\max{}}$ denote the maximum volatility among the $m$ stocks, we can show, as in Lemma~\ref{lemma-azuma}, that
\[
\Pr\big(T_n \le Ce^{-\sigma_{\max{}}\lambda\sqrt{m}} \text{ or } T_n \ge Ce^{\sigma_{\max{}}\lambda\sqrt{m}}\big) \le 2e^{-\lambda^2/2},
\]
for any $\lambda > 0$. The results of the single stock case are now applicable. Given $\varepsilon > 0$, choose $\lambda_0 = \sqrt{2\ln\frac{2}{\varepsilon}}$, so that $\Pr\big(T_n \le Ce^{-\sigma_{\max{}}\lambda\sqrt{m}} \text{ or } T_n \ge Ce^{\sigma_{\max{}}\lambda\sqrt{m}}\big) \le \varepsilon$. We sample $N$ paths $\omega^1, \ldots, \omega^N$ in $\Omega$ according to $\Pi$. Let $Z$ be be number of paths $\omega^i$ such that $T_n(\omega^i) \le (n+1)X$, where $X$ is the strike price of the basket call.

\begin{theorem}
Let $0 < \delta < 1$ be given. With $N = \Theta(\log\frac{1}{\delta} + \frac{1}{\varepsilon^2}e^{4\sigma_{\max{}}\lambda_0\sqrt{m}}\frac{1+ 2\sigma_{\max{}}\varepsilon\sqrt{m}}{\lambda_0 - 2\sigma_{\max{}}\sqrt{m}})$ trials, the following statements hold.
\begin{enumerate}
\item
If $\frac{Z}{N} \le 2\varepsilon$, then, with probability $1 - \delta$, $\E(T_n - (n+1)X)$ estimates $\E\big((T_n - (n+1)X)^+\big)$ with error at most $4\varepsilon(n+1)X$. Correspondingly, the price of the call is estimated with error at most $4\varepsilon X$.
\item
If $\frac{Z}{N} > 2\varepsilon$, then, with probability $1 - \delta$, $\frac{1}{N}\sum_{i=1}^N \big(T_n(\omega^i) - (n+1)X\big)^+$ estimates $\E\big((T_n - (n+1)X)^+\big)$ with standard deviation at most $\varepsilon(n+1)X$. Correspondingly, the price of the call is estimated with standard deviation at most $\varepsilon X$.
\end{enumerate}
\end{theorem}

\begin{proof}
This theorem is analogous to Theorem~\ref{theorem-algorithm} and can be proven using similar techniques.
\end{proof}

\section{A Recursive Bucketing-Based Scheme}
\label{section-aggregation}

The AMO algorithm takes $\Oh(kn^2)$ time to produce a price estimate in the range $\left[P - \frac{nX}{k}, P\right]$, where $P$ is the exact price of the call and $k$ is any natural number. As in Section~\ref{section-MC}, this algorithm estimates $\E\big((T_n - (n+1)X)^+\big)$, from which the price of the call, $\E((A_n - X)^+)$, can be easily estimated. Our recursive scheme is a generalization of a variant of the AMO algorithm. Below we first describe this variant, called \emph{Bucketed Tree Traversal} (\aggregate{}) and then detail our scheme, called \emph{Recursive} Bucketed Tree Traversal (\recAggregate{}).

Given binomial tree $\T$ of depth $n$, and numbers $t$, $i$, $m$ such that $0\le t\le n$, $0\le i\le t$, and $m \le n-t$, let $\T^{[t, i]}_m$ be the subtree of depth $m$ rooted at node $\T[t, i]$. Given $0\le t \le n$, let $\omega|_t$ be the prefix of $\omega$ up to level $t$ of $\T$. Note that $\omega|_n = \omega$. Given $\psi, \omega\in\Omega$, we say that $\psi$ is an \emph{extension} of $\omega|_m$ if, for $0\le t\le m$, we have $\psi_t = \omega_t$. Given another binomial tree $\U$ of depth $n$ we say that $\psi\in\Omega(\U)$ is \emph{isomorphic} to $\omega\in\Omega(\T)$, if, for all $0\le t\le n$, $\psi_t = \omega_t$.

Like the AMO algorithm, \aggregate{} is based on the following simple observation. Suppose that the running total $T_m(\omega) = T_m(\omega|_m)$ of the stock prices on path $\omega\in\Omega$ exceeds the \emph{barrier} $B = (n+1)X$. Then, for any extension $\psi$ of $\omega|_m$, $T_n(\psi)$ also exceeds $B$ and the call will be exercised. If we know the call will be exercised on all extensions of $\omega|_m$, it is easy to compute the payoff of the call on these extensions, as described in Lemma~\ref{lemma-expected}. 

As we travel down a path $\omega$, once the running total $T_m(\omega)$ exceeds $B$, we can just keep track of the running total on extensions $\psi$ of $\omega|_m$ weighted by $\Pi(\psi)$, from which the value of the option can be computed. Hence, we only need to individually keep track of path prefixes $\omega|_m$ that have running totals $T_m(\omega|_m)$ less than $B$.

Unfortunately, there may be exponentially many such $\omega|_m$. However, the running totals $T_m(\omega|_m)$ are in the bounded range $[0, B)$. Rather than trying to keep track of each running total individually, we instead group the running totals terminating at each node into \emph{buckets} that subdivide this interval. This introduces some round-off error. Suppose we use $k$ buckets to divide $[0, B)$ into equal-length subintervals and we use the left endpoint of each interval as the representative value of the running totals contained in that bucket. At each step down the tree, when we put a running total into a bucket, an error of at most $\frac{B}{k}$ is introduced. Traveling down $n$ levels of the tree, the total  $T_n(\omega)$ of a path $\omega$ is underestimated by at most $\frac{nB}{k}$ and the average $A_n(\omega)$ is underestimated by at most $\frac{B}{k}$.

\aggregate{} is detailed in Algorithm~\ref{algorithm-aggregate}. At each node $v = \T[t, i]$ of $\T$, create $k+1$ buckets to store partial sums of path prefixes terminating at $v$. There will be $k$ \emph{core buckets} and one \emph{overflow bucket}. This overflow bucket is the only difference between \aggregate{} and the AMO algorithm.  For $0\le j < k$, core bucket $b_j(v)$ stores the \emph{probability mass} $b_j(v)\mass$ of path prefixes that terminate at node $v$ and have running totals in its \emph{range} $\range(j) = [j\frac{B}{k}, (j+1) \frac{B}{k})$. The representative value of partial sums in the bucket is denoted by $b_j(v)\val = j\frac{B}{k}$. The overflow bucket $b_k(v)$ stores the probability-weighted running total estimates of path prefixes that have estimated running totals exceeding $B$. This quantity is denoted by $b_k(v)\val$. The probability mass of these path prefixes is denoted by $b_k(v)\mass$. \aggregate{} iterates through each of the $k+1$ buckets of each of the $\frac{(n+1)(n+2)}{2}$ nodes in $\T$, for a total runtime of $\Oh(kn^2)$.

\begin{algorithm}
\label{algorithm-aggregate}

\indentedline{0}\aggregate($\T$, $k$, $B$)
\indentedline{1}for each node $v\in\T$ and each bucket $b_j(v)$ 
\indentedline{2}set $b_j(v)\mass\leftarrow 0$;
\indentedline{1}take $j$ such that initial price $s(\T[0,0])\in \range(j)$; $b_j(\T[0,0])\mass\leftarrow 1$; \komment{assume $s(\T[0,0]) < B$}
\indentedline{1}for $t = 0,\ldots, (n-1)$ \komment{iterate through each level}
\indentedline{2}for $i = 0,\ldots, t$ \komment{iterate through each node at level $t$}
\indentedline{3}let $v = \T[t, i]$; \komment{shorthand notation for node $\T[t, i]$}
\indentedline{3}for $w\in\Set{\T[t+1, i], \T[t+1, i+1]}$ \komment{for each child of node $v$}
\indentedline{4}let $p'\in\{p,q\}$ be the probability of going from node $v$ to $w$;
\indentedline{4}for $b_j(v)\in\Set{b_0(v),\ldots,b_k(v)}$ \komment{for each bucket at node $v$}
\indentedline{5}let $V \leftarrow b_j(v)\val + s(w)$; 
\indentedline{5}let $M \leftarrow b_j(v)\mass\times p'$;
\indentedline{5}if $V < B$
\indentedline{6}take $\ell$ such that $V \in \range(\ell)$;
\indentedline{6}$b_\ell(w)\mass \leftarrow b_\ell(w)\mass + M$;
\indentedline{5}else\komment{in overflow bucket}
\indentedline{6}$b_k(w)\val \leftarrow \frac{b_k(w)\mass\times b_k(w)\val + M\times V}{b_k(w)\mass + M}$;
\indentedline{6}$b_k(w)\mass \leftarrow b_k(w)\mass + M$;
\indentedline{1}return $\sum_{i=0}^{n}b_k(n,i)\mass\times(b_k(n,i)\val - B)$.
\indentedline{1}\hfill\komment{return option price estimated from overflow buckets at leaves}
\end{algorithm}

We propose \recAggregate{,} a recursive extension of \aggregate{.} Consider some level $t$ in our binomial tree $\T$ and assume that the weights of all path prefixes terminating at level $t$ have been put into the appropriate buckets. \aggregate{} uses these weights to compute the bucket weights of nodes at level $t+1$. In contrast, \recAggregate{} recursively solves the problem for subtrees $\T^{[t, i]}_m$, $0\le i \le t$, of some depth $m < n-t$ rooted at node $\T[t, i]$.\footnote{Actually, the trees on which we recursively solve the problem are not exactly the $\T^{[t, i]}_m$. Each is identical to the respective $\T^{[t, i]}_m$, except the price at the root is changed from $s(\T[t, i])$ to $0$. The reason for this is explained in Remark~\ref{rootZero}.} As each recursive call is complete, \recAggregate{}  \merge{s} the bucket weights at the leaves of $\T^{[t, i]}_m$ into the corresponding nodes at level $t+m$ of $\T$. The advantages of the recursive calls are twofold. 
\begin{enumerate}
\item
They use finer bucket granularity, resulting in improved accuracy. 
\item
The results of a single recursive call on a particular subtree $\T^{[t, i]}_m$ are used to \estimate{} the results of other recursive calls to other subtrees $\T^{[t, j]}_m$, where $j>i$, as long as the node prices in $\T^{[t, j]}_m$ are ``sufficiently close'' to the corresponding node prices in $\T^{[t, i]}_m$. This improves the runtime, since we do not need to make all $t+1$ of the recursive calls, so there are portions of $\T$ that we do not directly traverse.
\end{enumerate}

\subsection{The \merge{} Procedure}

Consider a recursive call on the subtree $\T_1 = \T^{[t_0, i_0]}_{n_1}$ of depth $n_1$ rooted at node $v_0 = \T[t_0, i_0]$. A leaf $v_1 = \T_1[n_1,i_1]$ of $\T_1$ ($0\le i_1 \le n_1$) corresponds to the node $v_2 = \T[t_0+n_1, i_0+i_1]$ of $\T$. \merge{} incorporates the bucket weights at $v_1$ into the bucket weights at $v_2$. Recall that the recursive call on $\T_1$ is made with finer bucket granularity. Assume we use $k_1 = h_1k_0$ core buckets instead of just $k_0 = k$. We first combine each group of $h_1$ buckets at $v_1$ into a single bucket, so that we are left with $k_0$ buckets to merge into $v_2$. When we refer to a bucket $b_{j_1}(v_1)$ below, $0\le j_1 < k_0$, we mean one of these $k_0$ combined buckets.

The \merge{} procedure is described in Algorithm~\ref{algorithm-merge}. Consider first the core buckets. Let $0\le j_0, j_1, j_2 < k_0$ denote core bucket indices. Bucket $b_{j_0}(v_0)$ contains the mass of path prefixes in $\T$ terminating at $v_0$ whose running total estimates fall into the interval $\range(j_0) = [j_0\frac{B}{k_0}, (j_0+1)\frac{B}{k_0})$. Bucket 
$b_{j_1}(v_1)$ contains the mass of full paths in $\T_1$ terminating at node $v_1$ whose running total estimates fall into the interval $\range(j_1)$. This is equal to the mass of partial paths in $\T$ starting at node $v_0$ and terminating at node $v_2$ whose running total estimates fall into the interval $\range(j_1)$. Merging $\T_1$ into $\T$ involves merging each leaf $v_1$ of $\T_1$ into the corresponding node $v_2$ of $\T$. Once the merging procedure is done, $b_{j_2}(v_2)\mass$ is updated to contain the weight of path prefixes in $\T$ passing through node $v_0$ and terminating at node $v_2$ that have running total estimates in the interval $\range(j_2)$. The overflow buckets are handled similarly.

\begin{algorithm}
\label{algorithm-merge}
\indentedline{0}\merge($\T$, $\T_1 = \T^{[t_0, i_0]}_{n_1}$)
\indentedline{1}let $v_0 \leftarrow \T[t_0, i_0]$;
\indentedline{1}for $i_1 = 0,\ldots,n_1$ \komment{for each leaf of $\T_1$}
\indentedline{2}let $v_1 \leftarrow \T_1[n_1,i_1]$, $v_2 \leftarrow \T[t_0+n_1, i_0+i_1]$;
\indentedline{2}for $j_0 = 0,\ldots,k_0$ \komment{buckets in $v_0$}
\indentedline{3}for $j_1 = 0,\ldots,k_0$ \komment{buckets in $v_1$}
\indentedline{4}let $V \leftarrow b_{j_0}(v_0)\val + b_{j_1}(v_1)\val$;
\indentedline{4}let $M \leftarrow b_{j_0}(v_0)\mass\times b_{j_1}(v_1)\mass$;
\indentedline{4}if $V < B$
\indentedline{5}take $j_2$ such that $V\in\range(j_2)$;
\indentedline{5}$b_{j_2}(v_2)\mass\leftarrow b_{j_2}(v_2)\mass + M$;
\indentedline{4}else
\indentedline{5}$b_k(v_2)\val \leftarrow \frac{b_k(v_2)\mass\times b_k(v_2)\val + M\times V}{ b_k(v_2)\mass +M}$;
\indentedline{5}$ b_k(v_2)\mass \leftarrow b_k(v_2)\mass + M$.
\end{algorithm}

\begin{remark}
\label{rootZero}
Notice that the price at $v_0$ is counted twice: once in the path prefix from the root $\T[0,0]$ to $v_0$ and once in the partial path between $v_0$ and $v_2$. To address this issue, when we recursively solve the problem on the subtree $\T_1$, we set the price $s(\T_1[0,0])$ at the root to be $0$, ensuring that this price is counted once. This modification does not change our algorithms.
\end{remark}

\begin{lemma}
\label{lemma-merge-accuracy}
For an arbitrary node $v$, let $E(v)$ be the maximum amount by which running totals terminating at node $v$ are underestimated by the bucket values. Using the \merge{} algorithm, we have $E(v_2) \le E(v_0) + E(v_1) + \frac{B}{k}$.
\end{lemma}

\begin{proof}
Consider first the core buckets. Bucket $b_{j_0}(v_0)$ covers running totals in the range $[j_0\frac{B}{k}, (j_0+1)\frac{B}{k})$ and $b_{j_1}(v_1)$ covers running totals in the range $[(j_1)\frac{B}{k},  (j_1 + 1)\frac{B}{k})$. The concatenation of paths in these buckets covers running totals in the range $[(j_0 + j_1)\frac{B}{k}, (j_0 + j_1 + 2)\frac{B}{k})$. Assuming $j_0 + j_1 < k$, we put these paths in bucket $b_{j_0 + j_1}(v_2)$, and the running totals are underestimated by the original error of $E(v_0) + E(v_1)$, plus an additional $\frac{B}{k}$, since some partial paths that fall in the range of bucket $b_{j_0 + j_1 + 1}(v_2)$ are actually put in $b_{j_0 + j_1}(v_2)$. If $j_0 + j_1 \ge k$, then these paths are put in the overflow bucket. In this case, no additional error is introduced. The original error of $E(v_0) + E(v_1)$ is carried over. Similar arguments can be made when we are merging overflow buckets.
\end{proof}

\begin{lemma}
\label{lemma-merge-runningtime}
\merge{} can be made to run in $\Oh(n_1k\log k)$ time.
\end{lemma}

\begin{proof}
In the implementation of Algorithm~\ref{algorithm-merge}, \merge{} runs in $\Oh(n_1k^2)$ time. However, the core buckets can be merged with a faster technique. Let $a_0(x) = \sum_{j_0=0}^{k-1} b_{j_0}(v_0)\mass \cdot x^{j_0}$ and $a_1(x) = \sum_{j_1=0}^{k-1} b_{j_1}(v_1)\mass \cdot x^{j_1}$ be polynomial representations of the bucket masses at nodes $v_0$ and $v_1$. Let $a_2(x) = a_0(x)\cdot a_1(x)$. This product can be computed in $\Oh(k\log k)$ time with the Fast Fourier Transform (FFT) \cite{Cormen:1999:IA}. For $0\le j_2 < k$, the coefficient of the $x^{j_2}$ term of $a_2(x)$ is the probability mass that should be added to bucket $b_{j_2}(v_2)$. For $j_2 > k$, this coefficient should be added to the overflow bucket $b_k(v_2)$. So the core buckets can be merged in $\Oh(k\log k)$ time per node. Merging the overflow buckets takes $\Oh(k)$ time, so the total runtime per node is $\Oh(k\log k)$.
\end{proof}

\subsection{The \estimate{} Procedure}

Let $\T_1 = \T^{[t_0, i_1]}_{m}$ and $\T_2 = \T^{[t_0, i_2]}_{m}$ be two subtrees of $\T$, where $i_2 > i_1$ We now describe the \estimate{$(\T_2, \T_1)$} procedure, which estimates the weights in the leaf buckets of $\T_2$ from the weights in the leaf buckets of $\T_1$. This saves us the work of recursively solving the problem on $\T_2$.

\estimate{} is described in Algorithm~\ref{algorithm-estimate}. It uses the following fact. Given any node $v_1 = \T_1[t, i]$ in $\T_1$, let $v_2 = \T_2[t, i]$ be the corresponding node in $\T_2$. Notice that there is a constant $\alpha > 1$ such that for all $(v_1, v_2)$ pairs, $s(v_2) = \alpha s(v_1)$. Hence, for any path $\psi\in\Omega(\T_2)$, we have $T_{m}(\psi) = \alpha T_{m}(\omega)$, where $\omega\in\Omega(\T_1)$ is isomorphic to $\psi$.

\begin{algorithm}
\label{algorithm-estimate}
\indentedline{0}\estimate($\T_2 = \T^{[t_0, i_2]}_{m}$, $\T_1 = \T^{[t_0, i_1]}_{m}$)
\indentedline{1}for $i = 0,\ldots, m$ \komment{go through the leaf buckets of $\T_1$ and $\T_2$}
\indentedline{2}let $v_1 \leftarrow \T_1[m, i]$, $v_2 \leftarrow \T_2[m, i]$;
\indentedline{2}for $j_1 = 0,\ldots, k$ \komment{go through each bucket at $v_1$}
\indentedline{3}let $V \leftarrow \alpha b_{j_1}(v_1)\val$;
\indentedline{3}let $M \leftarrow b_{j_1}(v_1)\mass$;
\indentedline{3}if $V < B$
\indentedline{4}take $j_2$ such that $V\in\range(j_2)$;
\indentedline{4}$b_{j_2}(v_2)\mass\leftarrow b_{j_2}(v_2)\mass + M$;
\indentedline{3}else
\indentedline{4}$b_k(v_2)\val \leftarrow \frac{b_k(v_2)\mass\times b_k(v_2)\val + M\times V}{ b_k(v_2)\mass +M}$;
\indentedline{4}$ b_k(v_2)\mass \leftarrow b_k(v_2)\mass + M$;
\end{algorithm}

\begin{lemma}
\label{lemma-estimate-accuracy}
Suppose that $\alpha \le 2$ and assume that the total path sums in $\T_1$ are underestimated by our bucketing scheme by at most $E$. \estimate{} underestimates the total path sums in $\T_2$ by at most $2E + 2\frac{B}{k}$.
\end{lemma}

\begin{proof}
Take any path $\omega\in\Omega(\T_1)$ and let $U_m(\omega)$ be the estimate for  $T_m(\omega)$ made by the bucketing scheme. By assumption, $T_m(\omega) - U_m(\omega) \le E$. Let $\psi\in\Omega(\T_2)$ be isomorphic to $\omega$. Using $\alpha U_m(\omega)$ as an estimate for $T_m(\psi) = \alpha T_m(\omega)$, we underestimate $T_m(\psi)$ by at most $\alpha E \le 2E$. This accounts for the first error term.

Now, consider core bucket $b_{j_1}(v_1)$ of some leaf $v_1$ of $\T_1$. The range of running totals covered by this bucket is $\range(j_1) = [j_1\frac{B}{k}, (j_1+1)\frac{B}{k})$. When we map this bucket to $\T_2$, it covers path sums in the range $[\alpha j_1\frac{B}{k}, \alpha (j_1+1)\frac{B}{k})$, which falls into at most three buckets in the corresponding node $v_2$ of $\T_2$, since $\alpha \le 2$. Call these buckets $b_{j_2}(v_2)$, $b_{j_2+1}(v_2)$, and $b_{j_2 + 2}(v_2)$ (assume these are all core buckets). Putting the entire mass into $b_{j_2}(v_2)$ causes the portion of partial sums that fall into bucket $b_{j_2 + 2}(v_2)$ to be underestimated by additional $2\frac{B}{k}$, which accounts for the second error term.

Similar errors occur if the range $[\alpha j_1\frac{B}{k}, \alpha (j_1+1)\frac{B}{k})$ spills into the overflow bucket $b_k(v_2)$ and when mapping the overflow bucket $b_k(v_1)$ in $v_1$ to $\T_2$.
\end{proof}

\begin{lemma}
\label{lemma-estimate-runtime}
Suppose that we would like to determine the leaf bucket weights of the subtrees $\T^{[t, i]}_m$, where $0\le i \le t$. We need only call \recAggregate{} at most once for every $\Theta(\frac{\sqrt{n}}{\sigma})$ subtrees, and use the \estimate{} procedure to estimate the leaf bucket weights of the other subtrees with bounded error.
\end{lemma}

\begin{proof}
We begin by recursively solving the problem on the subtree $\T^{[t, 0]}_m$. Notice that $s(\T[t, i]) = u^{2i}s(\T[t, 0])$. By Lemma~\ref{lemma-estimate-accuracy}, as long as $u^{2i} \le 2$, we can use \estimate{} to compute the leaf bucket weights of $\T^{[t, i]}_m$, without introducing significant error. Recalling that we take $u = e^{\sigma/\sqrt{n}}$, where $\sigma$ is the volatility of the stock, we see that we can use \estimate{$(\T^{[t, i]}_m, \T^{[t, 0]}_m)$} for $i$ as large as $\frac{\sqrt{n}\ln 2}{2\sigma} = \Theta(\frac{\sqrt{n}}{\sigma})$ while keeping the error bounded. This proves the lemma.
\end{proof}

\subsection{Error and Runtime Analysis}
\label{subsection-recAggregateAnalysis}

We now derive recursive expressions for the error and runtime of \recAggregate{.} Suppose there are $n_0$ trading periods and we use a total of $k_0$ buckets per node. For $i> 0$, let $n_i$ be the number of trading periods and $k_i$ the number of buckets per node we use in the $i$-th subproblem (at the $i$-th level into the recursion). Let $E_i = E(n_i, k_i)$ be the \emph{error} bound made in the $i$-th subproblem. The error is the amount by which \recAggregate{} underestimates the running totals. Let $T_i = T(n_i, k_i)$ be the runtime of \recAggregate{} on the $i$-th subproblem, including all recursive calls. We recursively call \recAggregate{} on binomial trees of decreasing depth (decreasing $n_i$) with an increasing number of buckets (increasing $k_i$) until we reach a tree of some small depth.

\begin{lemma}
\label{lemma-error}
When $i$ recursive calls are made, the error made by \recAggregate{} is
\[
E_0 = 5Bn_0\sum_{j=0}^i \frac{2^j}{k_jn_{j+1}} + \frac{2^{i+1}n_0}{n_{i+1}}\tilde{E}_{i+1},
\]
where $\tilde{E}_{i+1}$ is the error in solving the $(i+1)$-st subproblem.
\end{lemma}
\begin{proof}
Consider the error in the $i$-th subproblem, $E_i = E(n_i, k_i)$. Each recursive call has an error of $E_{i+1}$. When the $k_{i+1}$ smaller granularity buckets in the $(i+1)$-st problem are grouped into the $k_i < k_{i+1}$ buckets of the $i$-th problem, an error of one bucket size, $\frac{B}{k_i}$, is introduced. A call to \estimate{} underestimates the price sums by at most $2(E_{i+1} + \frac{B}{k_i}) + 2\frac{B}{k_i} = 4\frac{B}{k_i} + 2E_{i+1}$, by Lemma~\ref{lemma-estimate-accuracy}. This error is made at most $\frac{n_i}{n_{i+1}}$ times, once at each of levels $n_{i+1}, 2n_{i+1}, \ldots, \frac{n_i}{n_{i+1}}n_{i+1}$. At each of these levels, an additional error of $\frac{B}{k_i}$ is introduced by the \merge{} procedure, as described in Lemma~\ref{lemma-merge-accuracy}. The total error in the $i$-th subproblem is
\[
E_i = \frac{5 B n_i}{k_i n_{i+1}} + \frac{2 n_i}{n_{i+1}}E_{i+1}. 
\]
Unraveling $i$ levels of the recursion on the original problem yields the result.
\end{proof}

For the runtime of \recAggregate{,} we must analyze how many times the $(i+1)$-st subproblem needs to be solved when solving the $i$-th subproblem. At each of levels $0, n_{i+1}, \ldots, (\frac{n_i}{n_{i+1}} - 1)n_{i+1}$ in the binomial subtree of the $i$-th subproblem, suppose the $(i+1)$-st subproblem is called at most $c_i$ times.

\begin{lemma}
\label{lemma-runtime}
When $i$ recursive calls are made, the runtime of \recAggregate{} is
\[
T_0 = \Oh\left(\sum_{j=0}^i n_0 n_j k_j\log(k_j)\prod_{k=0}^{j-1}c_k\right) + \frac{n_0\prod_{k=0}^i c_k}{n_{i+1}}\tilde{T}_{i+1},
\]
where $\tilde{T}_{i+1}$ is the time it takes to solve the $(i+1)$-st subproblem.
\end{lemma}

\begin{proof}
Consider the runtime of the $i$-th subproblem, $T_i = T(n_i, k_i)$. Each recursive call takes time $T_{i+1}$ and at most $c_i$ recursive calls are made at each of levels $0, n_{i+1}, \ldots, (\frac{n_i}{n_{i+1}} - 1)n_{i+1}$. The leaf bucket weights of the other at most $n_i$ subtrees rooted at the nodes at each of these levels can be determined using \estimate{,} which takes $\Oh(k_i n_{i+1})$ time. For the \merge{} procedure, at each of $\frac{n_i}{n_{i+1}}$ levels $t$, each of the at most $n_i$ nodes at level $t$ must be merged with $n_{i+1}$ leaves from the subtrees $\T^{[t, j]}_{n_{i+1}}$ to get the bucket weights at level $t+n_{i+1}$. The merging of each node takes $\Oh(k_i\log k_i)$ time. The total runtime of the $i$-th subproblem is
\[
\fl
T_i = \frac{n_i}{n_{i+1}}
(c_i T_{i+1} + \Oh(n_i k_i n_{i+1}) + \Oh(n_i n_{i+1} k_i\log k_i)) = 
\frac{n_i c_i}{n_{i+1}} T_{i+1} + \Oh(n_i^2 k_i\log k_i).
\]
Unraveling $i$ levels of the recursion on the original problem yields the lemma.
\end{proof}

\begin{theorem}
\label{theorem-recAggregate}
Given integer $R>2$, let $\gamma = \frac{1}{R}$, and for $i>0$, let $n_i = (\frac{n_0}{\sigma^2})^{1/2 - i\gamma}$ and $k_i = 4^ik_0(\frac{n_0}{\sigma^2})^{i\gamma}$, where $\sigma$ is the volatility of the stock. \recAggregate{} underestimates $\E\big((T_n - (n+1)X)^+\big)$ by at most $\Oh\big(\frac{Bn_0^{1/2 +\gamma}\sigma^{1 - 2\gamma}}{k_0}\big)$ and takes time 
$
\Oh\big(2^{1/\gamma}n_0^2k_0(\frac{1}{\gamma} + \log \frac{k_0n_0}{\sigma^2})\big).
$
\end{theorem}

\begin{proof}
After $K = \frac{1}{2\gamma}$ levels of recursion, we have $n_K = (\frac{n_0}{\sigma^2})^0 = 1$. The problem consists of a single node and can be solved in constant time $\tilde{T}_{K}$ with an error of $\tilde{E}_K = \frac{B}{k_K} = \frac{\sigma B}{4^{K}k_0 n_0^{1/2}}$. >From Lemma~\ref{lemma-error}, the error is
\begin{eqnarray*}
\fl
E_0 = 5Bn_0\sum_{i=1}^K\frac{2^{i-1}}{k_{i-1}n_i} + \frac{2^Kn_0}{n_K}\tilde{E}_K
= 5Bn_0\sum_{i=1}^K\frac{1}{2^{i-1}k_0(\frac{n_0}{\sigma^2})^{1/2 - \gamma}} + \frac{2^Kn_0\sigma B}{4^Kk_0n_0^{1/2}} \\
\lo= \Oh\left(\frac{\sigma^{1 - 2\gamma}Bn_0^{1/2 + \gamma}}{k_0}\right).
\end{eqnarray*}
For the runtime analysis, by Lemma~\ref{lemma-estimate-runtime}, note that $c_0 = \frac{n}{\sqrt{n}/\sigma}= \sigma\sqrt{n}$. For $i>0$, since $n_i < \frac{\sqrt{n}}{\sigma}$, $c_i = 1$. Combining this fact with Lemma~\ref{lemma-runtime} yields the runtime
\begin{eqnarray*}
T_0 &=& n_0^2 k_0\log k_0 + \Oh\left(\sum_{i=1}^{K-1}n_0n_ic_0k_i\log k_i\right) + \Oh(n_0c_0) \\
&=& n_0^2 k_0\log k_0 + \Oh\left(\sum_{i=1}^{K-1}n_0^2 4^i k_0 (\textstyle\frac{1}{\gamma} + \log \textstyle\frac{k_0 n_0}{\sigma^2})\right) \\
&=& \Oh(2^{1/\gamma} n_0^2 k_0 (\textstyle\frac{1}{\gamma} + \log \textstyle\frac{k_0 n_0}{\sigma^2})).
\end{eqnarray*}
\end{proof}

\begin{corollary}
Given integer $R>2$, let $\gamma = \frac{1}{R}$ and choose $n_i$ and $k_i$ as in Theorem~\ref{theorem-recAggregate}. In time 
$
\Oh\big(2^{1/\gamma}n_0^2k_0(\frac{1}{\gamma} + \log \frac{k_0n_0}{\sigma^2})\big),
$
\recAggregate{} returns an option price in the range $\big[P - \frac{X n_0^{1/2 + \gamma}\sigma^{1 - 2\gamma}}{k_0}, P\big]$, where $P$ is the exact price of the option.
\end{corollary}

\begin{proof}
This follows from Theorem~\ref{theorem-recAggregate} and the fact that 
\[
\E\big((A_n - (n+1)X)^+\big) = \frac{1}{n+1}\E\big((T_n - (n+1)X)^+\big).
\]
\end{proof}

\begin{remark}
By comparison, the AMO algorithm \cite{Aingworth:2000:AAA} has error $\Oh(n_0X/k_0)$ and
runtime $\Oh(k_0n_0^2)$.
\end{remark}

\subsection{Customization of the \recAggregate{} Scheme}
\label{sec_custom}

A useful feature of the \recAggregate{} scheme is that the number of recursive calls, the size of each recursive call, and the pricing method by which we solve the base case of the recursion (the final recursive call) can be custom tailored for the application. The base case of the recursion can be solved with any option pricing scheme, including \strongMC{}, other variants of the Monte Carlo algorithm, the AMO algorithm, or exhaustive path traversal. For example, consider a 3-month (90-day) Asian option with the average computed on a daily basis ($n=90$). Pricing this option exactly requires traversing $2^{90}$ paths, which is infeasible. Using our \recAggregate{} scheme, since $\sqrt{n} < 10$, a single recursive call can reduce this problem to several subproblems each with at most $2^{10}$ paths.

In practice, larger values of $n$ appear in several applications. Long-term options contracts, LEAPS \cite{Hull:2000:OFO}, are negotiated for exercise dates several years in advance. Companies also offer stock options to employees over periods of up to five years or more. Finally, in \emph{Istanbul} options contracts \cite{Jacques:1997:IOW}, rather than computing the average based on daily prices, the average is computed over prices taken at higher frequencies. In each of these cases, $n$ is sufficiently large that several recursive calls are required to reduce the problem to a manageable size.

\section{Pricing Basket Options}
\label{section-basket}

The bucketed tree structure created by \aggregate{} and \recAggregate{} can be used to price various kinds of European basket options as well. Here, we describe here how to price European Asian basket options. A basket option \cite{Hull:2000:OFO} is composed of $m$ stocks, $z_1, \ldots, z_m$. For each stock $z_i$, we construct a binomial tree (according to the respective stock's volatility), as described in Section~\ref{section-definitions}. For $1\le t\le n$ and $1\le i\le m$, let $S_t^i$ be the random variable denoting the price of stock $z_i$ on day $t$ and define $S_t = \sum_{i=1}^m S_t^i$ to be the random variable denoting the total of the stock prices on day $t$. Recall that the payoff of a European basket call with strike price $X$ is $\E((S_n - X)^+)$. Letting $A_n = \frac{1}{n+1}\sum_{t=0}^n S_t$ be the average total stock price, the payoff of a European Asian basket call with strike price $X$ is $\E((A_n - X)^+)$. There is additional complexity with pricing basket options that does not appear in their single-stock counterparts: the number of paths that the total basket price can follow is exponential, not only in the number of trading periods $n$ but also in that of stocks $m$. Basket options are usually priced using traditional Monte Carlo methods. The scheme we describe here is the first polynomial time, in both the number of stocks and trading periods, pricing scheme for any kind of basket option with provably small error bounds.

Our European Asian basket call pricing algorithm, \basketAggregate{,} is described in Algoirthm~\ref{algorithm-basket}. Let $B = (n+1)X$, where $X$ is the strike price of the basket option. For each stock $z_i$, $1\le i\le m$, use \recAggregate{} to construct the bucketed binomial tree structure $\T^i$ described in Section~\ref{section-aggregation}, this time using $B$ as the barrier; should the running total of any $z_i$ exceed $B$, the basket option will always be exercised, regardless of what the other stocks do. For each stock $z_i$, we construct $k+1$ \emph{superbuckets} $\beta^i_j$, $0\le j\le k$, where $\beta^i_j$ is the combination of buckets $b_j(v)$ for all leaves $v\in\T^i$. For the core buckets $\beta_j^i$, $0\le j < k$, let $\beta^i_j\val = j\frac{B}{k}$ and $\beta^i_j\mass = \sum_{\ell = 0}^n b_j(\T^i[n, \ell])\mass$, where this summation ranges over all leaves $\T^i[n, \ell]$ of $\T^i$. For the overflow bucket $\beta_k^i$, let $\beta^i_k\mass = \sum_{\ell = 0}^n b_j(\T^i[n, \ell])\mass$ and 
\[
\beta^i_k\val = \frac{\sum_{\ell = 0}^n b_k(\T^i[n, \ell])\val \times b_j(\T^i[n, \ell])\mass}{\beta^i_k\mass}.
\]

\paragraph{Handling overflow superbuckets}
If the running total of a stock $z_i$ reaches the overflow superbucket $\beta^i_k$, the option will be exercised regardless of what the other stocks do. Given this, we can determine the value of the option exactly, since 
\begin{eqnarray*}
\E\big((T_n - (n+1)X)^+\big) = \E\big(T_n - (n+1)X\big) = \E(T_n) - (n+1)X \\
= \beta^i_k\val + \sum_{i'\not= i}\E(T_n^{i'}) - (n+1)X,
\end{eqnarray*}
where $T_n^{i'}$ is the random variable denoting the running total of stock $z_{i'}$ up to day $n$. $\E(T_n^{i'})$ can be computed exactly using Lemma~\ref{lemma-expected}.

\paragraph{Handling core superbuckets}
Consider now the core superbuckets $\beta^i_j$, $0\le j < k$. Let $f_i(x) = \sum_{j=0}^{k-1} \beta^i_j\mass\cdot x^{j}$ be the polynomial representation of the core bucket masses of stock $z_i$ and let $f(x) = \prod_{i=1}^m f_i(x)$. This product can be computed efficiently, as described in Lemma~\ref{lemma-basket-runtime}. Notice that $f(x)$ has the form 
\[
f(x) = b_0 x^{0} + b_1 x^{1} + \cdots + b_{m(k-1)} x^{m(k-1)}. 
\]
>From the definition of $f(x)$, observe that $b_j$ is just the probability that the sum (over all stocks $z_i$) of running totals $T^i_n$ from the core buckets falls in the range $[j\frac{B}{k}, (j+1)\frac{B}{k})$. That is, 
\[
b_j = \Pr\left(\sum_{i=1}^m T^i_n \in [j\textstyle\frac{B}{k}, (j+1)\textstyle\frac{B}{k}) \ | \ \text{$T^i_n < B$ for all $i$}\right).
\]
Hence, the contribution to the option price from the core buckets can be estimated by
\[
\sum_{j=k}^{m(k-1)} b_j (j\textstyle\frac{B}{k} - (n+1)X).
\]

\paragraph{Pricing the option}
Combining the above results for the overflow and the core superbuckets, we see that $\E\big((T_n - (n+1)X)^+\big)$ can be estimated by 
\[
\fl
\sum_{i=1}^m\beta^i_k\mass\big(\beta^i_k\val + {\textstyle\sum_{i'\not= i}\E(T_n^{i'})} - (n+1)X\big) + \sum_{j=k}^{m(k-1)} b_j \big(j\textstyle\frac{B}{k} - (n+1)X\big).
\]

\begin{algorithm} 
\label{algorithm-basket}
\basketAggregate{}$(z_1, \ldots, z_m, B = (n+1)X)$
\indentedline{1} for $i = 1,\ldots, n$ \komment{for each stock $z_i$}
\indentedline{2} compute $\E(T^i_n)$ using Lemma~\ref{lemma-expected};
\indentedline{2} run \recAggregate{} on stock $z_i$ with barrier $B$ and $k$ buckets;
\indentedline{2} for $j = 0, \ldots, k-1$ \komment{construct core superbuckets}
\indentedline{3} $\beta^i_j\val \leftarrow j\frac{B}{k}$;
\indentedline{3} $\beta^i_j\mass \leftarrow \sum_{\ell = 0}^n b_j(\T^i[n, \ell])\mass$;
\indentedline{2} $\beta^i_k\mass \leftarrow \sum_{\ell = 0}^n b_j(\T^i[n, \ell])\mass$; \komment{construct overflow superbucket}
\indentedline{2} $\beta^i_k\val \leftarrow \frac{1}{\beta^i_k\mass}\sum_{\ell = 0}^n b_k(\T^i[n, \ell])\val \times b_j(\T^i[n, \ell])\mass$;
\indentedline{2} let $f_i(x) \leftarrow \sum_{j=0}^{k-1} \beta^i_j\mass\ x^{j}$;
\indentedline{1} compute $f(x) \leftarrow \prod_{i=1}^m f_i(x)$ as described in Lemma~\ref{lemma-basket-runtime};
\indentedline{1} let $f(x) \leftarrow  b_0 x^0 + b_1 x^1 + \cdots + b_{m(k-1)} x^{m(k-1)}$; \komment{for some $b_0, \ldots, b_{m(k-1)}$}
\indentedline{1} return $\sum_{i=1}^m\beta^i_k\mass\big(\beta^i_k\val + \sum_{i'\not= i}\E(T_n^{i'}) - (n+1)X\big) + \sum_{j=k}^{m(k-1)} b_j \big(j\textstyle\frac{B}{k} - (n+1)X\big)$;
\end{algorithm}

\begin{lemma}
\label{lemma-basket-runtime}
Let $\sigma_{\min{}}$ be the minimum volatility among the stocks in the basket. The runtime of \basketAggregate{} is 
\[
\Oh\big(mT(\sigma_{\min{}}, n,k) + mk\log m\log k + mk\log^2m\big),
\]
where $T(\sigma_{\min{}}, n,k)$ is the runtime of \recAggregate{} on a binomial tree of size $n$ when $k$ buckets are used and the volatility of the underlying stock is $\sigma_{\min{}}$.
\end{lemma}

\begin{proof}
Computing the leaf bucket weights for each stock $z_i$ (using \recAggregate{}) takes $\Oh(m\cdot T(\sigma_{\min{}}, n,k))$ time. The runtime of the rest of the computation is dominated by the time to compute $\prod_{i=1}^m f_i(x)$. We now describe an efficient way to do this. Assume that $m=2^r$, for some $r$. The general case is handled similarly. We conduct the multiplication by successively multiplying consecutive polynomials together $\log m = r$ times until we are left with a single polynomial. For $1\le i\le m$, let $f^0_i(x) = f_i(x)$. For $1\le j\le r$ and $1\le i\le \frac{m}{2^i}$, let $f^j_i(x) = f^{j-1}_{2i-1}(x)f^{j-1}_{2i}(x)$. The answer that we are looking for is just $\prod_{i=1}^m f_i(x) = f^r_1(x)$. At stage $j$, we multiply together $\frac{m}{2^j}$ pairs of polynomials, each of degree $2^{j-1}k$. Using FFT, this takes $\Oh\big(\frac{m}{2^j}2^{j-1}k\log(2^{j-1}k)\big) = \Oh\big(mk\log(2^{j-1}k)\big)$ time. The total runtime to compute the product, over all stages, is $\Oh\big(mk(r\log k + \sum_{j=1}^r\log2^{j-1})\big) = \Oh\big(mk(\log m\log k + \log^2m)\big)$, from which the claimed result follows.
\end{proof}

Our definition of the \emph{error} made by \basketAggregate{} is symmetric to the definition of error made by \recAggregate{;} \ie{,} it is the maximum amount by which \basketAggregate{} can underestimate $\sum_{i=1}^m \sum_{t=0}^n S_t^i(\omega^i)$, for paths $\omega^i\in\T^i$.

\begin{lemma}
\label{lemma-basket-error}
Let $\sigma_{\max{}}$ be the maximum volatility among the stocks in the basket.
The error made by \basketAggregate{} is at most $mE(\sigma_{\max{}}, n,k)$, where $E(\sigma_{\max{}}, n,k)$ is the error made by \recAggregate{} on a single stock with volatility $\sigma_{\max{}}$.
\end{lemma}

\begin{proof}
For $1\le i\le m$, let $\omega_i$ be any path down the binomial tree corresponding to stock $z_i$ and let $T^i_n(\omega_i)$ be the total price down $\omega_i$. When we run \recAggregate{} on $z_i$, each $T^i_n(\omega_i)$ is underestimated by at most $E(\sigma_{\max{}}, n,k)$. Hence for any $\omega_1,\ldots,\omega_m$, $\sum_{i=1}^m T^i_n(\omega_i)$ is underestimated by at most $mE(\sigma_{\max{}}, n,k)$, as claimed.
\end{proof}

\begin{theorem}
\label{theorem-basket}
Given $n$, $m$, $k$, $R>2$, $\gamma = \frac{1}{R}$, $\sigma_{\min{}}$ and $\sigma_{\max{}}$, if we apply \recAggregate{} as described in Theorem~\ref{theorem-recAggregate} to construct the bucketed binomial tree for each stock, \basketAggregate{} has an error of $\Oh(m\frac{Bn^{1/2 + \gamma}\sigma_{\max{}}^{1 - 2\gamma}}{k})$ and runs in time 
\[\textstyle
\Oh\big(2^{1/\gamma}n^2mk(\frac{1}{\gamma} + \log \frac{kn}{\sigma_{\min{}}^2}) + mk\log m\log k + mk\log^2 m\big).
\]
\end{theorem}

\begin{proof}
This follows directly from Lemmas~\ref{lemma-basket-runtime} and \ref{lemma-basket-error} and Theorem~\ref{theorem-recAggregate}.
\end{proof}

\begin{corollary}
Given $n$, $m$, $k$, $R>2$, $\gamma = \frac{1}{R}$, $\sigma_{\min{}}$ and $\sigma_{\max{}}$, \basketAggregate{} underestimates the price of a European Asian basket call by at most $\Oh(m\frac{Xn^{1/2 + \gamma}\sigma_{\max{}}^{1 - 2\gamma}}{k})$ and runs in time 
\[\textstyle
\Oh\big(2^{1/\gamma}n^2mk(\frac{1}{\gamma} + \log \frac{kn}{\sigma_{\min{}}^2}) + mk\log m\log k + mk\log^2 m\big).
\]
\end{corollary}

\begin{proof}
This follows from Theorem~\ref{theorem-basket}, which gives the amount by which the total prices of the stocks may be underestimated. The amount by which the sum of their daily averages is underestimated is smaller by a factor of $\frac{1}{n+1}$.
\end{proof}

\section{Further Research}
\label{section-open}

This paper focuses on theoretical aspects of \strongMC{} and
\recAggregate{}.  In order to optimize the performance of
\recAggregate{} in practice, we must consider three issues in its
implementation: (1) The number of recursive calls and the size of the
recursive subproblems will depend on the number of periods (see
Section~\ref{sec_custom}). (2) For options with a high number of
periods, nontrivial data structures may be required.  (3) We have a
choice of several algorithms to use at the base-case of our recursion
(see Section~\ref{sec_custom}). The implementation of our algorithms
and empirical comparisons between them and others will be reported in
a subsequent paper.

This paper opens up several directions for further theoretical
research. An immediate open problem is whether similar Monte Carlo and
bucketing techniques can be used for pricing American Asian
single-stock and basket options.

With regards to our algorithms, it would be an important result if the
runtime of \strongMC{} on basket options can be bounded by a
polynomial in the number of stocks $m$; or if the recursive calls in
\recAggregate{} can be structured differently to yield improved
runtime or error bounds.

Finally, it would be interesting to see whether our Monte Carlo
analysis or recursive bucketing techniques can be used to price other
types of options or applied to similar problems.

\ack

We would like to thank Kyusik Chung, David Goldenberg, and Samuel Ieong for helpful discussions; and Stan Eisenstat for pointers to numerical analysis software.

\section*{References}

\bibliographystyle{abbrv}
\bibliography{paper}

\end{document}